\documentclass[12pt]{article}
\usepackage{graphicx}

\begin{document}

\baselineskip 0.1667in

\begin{center}
{\large \textbf{A Proposed Mechanism for the Intrinsic Redshift}}

{\large \textbf{and its Preferred Values Purportedly Found in Quasars}}

{\large \textbf{Based on the Local-Ether Theory}}

\vspace{1cm}

\textsf{Ching-Chuan Su}

Department of Electrical Engineering

National Tsinghua University

Hsinchu, Taiwan

\vspace{1cm}
\end{center}

\noindent \textbf{Abstract}\textit{\ }-- Quasars of high redshift may be
ejected from a nearby active galaxy of low redshift. This physical
association then leads to the suggestion that the redshifts of quasars are
not really an indication of their distances. In this investigation, it is
argued that the high redshift can be due to the gravitational redshift as an
intrinsic redshift. Based on the proposed local-ether theory, this intrinsic
redshift is determined solely by the gravitational potential associated
specifically with the celestial object in which the emitting sources are
placed. During the process with which quasars evolve into ordinary galaxies,
the fragmentation of quasars and the formation of stars occur and hence the
masses of quasars decrease. Thus their gravitational potentials and hence
redshifts become smaller and smaller. This is in accord with the aging of
redshift during the evolution process. In some observations, the redshifts
of quasars have been found to follow the Karlsson formula to exhibit a
series of preferred peaks in their distributions. Based on the quasar
fragmentation and the local-ether theory, a new formula is presented to
interpret the preferred peaks quantitatively.

\vspace{0.5cm}

\noindent Subject headings: quasars: general --- galaxies: distances and
redshifts --- galaxies: active

\vspace{1.5cm}

\noindent {\large \textbf{1. Introduction}}\\[0.2cm]
Quasars are known for their high redshifts. It is widely accepted in
astronomy that the redshift is due to the Doppler effect which in turn
depends on the receding speed of the emitting source with respect to the
observer. According to the Hubble law, the higher the redshift, the farther
the source is away from the observer. Thereby, quasars are expected to move
quite fast and be located quite far away from the Earth. Meanwhile, several
observations seem to indicate that some quasars of high redshifts are quite
close in angular position to an active galaxy of low redshift and high
luminosity. This closeness is generally expected to be merely a projection
effect as a fortuitous coincidence in their angular positions. However, some
phenomena with morphological connections have been further observed. That
is, quasars occur in a pair and are located on opposite sides of the axis of
rotation of the active center galaxy. Further, the pair of quasars can be of
similar redshifts and distances from the galaxy. Moreover, also along the
axis there can be some other pairs of quasars or galaxies located farther
from the center galaxy. There is a trend that the center galaxy is of the
lowest redshift and the strongest luminosity, the nearby quasar is of the
highest redshift and the weakest luminosity, and of the companion quasars or
galaxies the redshifts decrease and the luminosities increase with their
distances away from the center galaxy (Arp \& Russell 2001). Thus the high
redshifts of quasars may not necessarily represent their distances. Further,
in some observations, the redshifts have been found to exhibit some
periodicity in their distributions as represented by the Karlsson formula
(Arp et al. 1990, Burbidge \& Napier 2001). The periodicity further makes it
difficult for the redshift to represent distance.

Thus it may be expected that the quasars are physically associated with the
center galaxy. Specifically, an ejection model has been proposed to
interpret the physical association. Thereby, a plausible mechanism is that
the center galaxy is an active one and ejects gas out of it (Arp \& Russell
2001). Then a nearby quasar is formed from the cloud of gas associated with
the ejection. By virtue of the conservation of momentum with the ejection,
quasars tend to form in pairs on opposite sides of the active galaxy. Then
quasars may fragment into pieces of smaller sizes, while they gradually
condense due to gravitation. Thus quasars evolve eventually into ordinary
galaxies by quasar fragmentation and star formation while they move away
from the parent galaxy. Meanwhile, further ejection forms a newborn pair of
quasars close to the parent galaxy. In this mechanism, the difference in
redshift can be related to the evolution of quasars and galaxies.

If the quasars are physically associated with the nearby parent active
galaxy, then the difference in redshift cannot be entirely attributed to the
Doppler effect. Accordingly, the redshift, at least part of it, should be
due to some kind of intrinsic properties. One model proposed for the
intrinsic redshift is the time-varying mass. That is, the masses of the
fundamental particles which make up the atoms are very small for newly
created matter, and then they increase with time (Arp \& Russell 2001). As
transition frequencies of atoms depend on the mass, the optical spectra will
shift as the mass is varying. Alternatively, it has been pointed out earlier
that the redshifts of quasars may be gravitational in origin and they are
likely to be very massive objects (Burbidge \& Burbidge 1967). Thus the
intrinsic redshift is attributed to the gravitational redshift which in turn
has been observed in the earthbound Pound-Rebka experiment dealing with the
emission and absorption of gamma-ray by employing iron crystals placed at
two different altitudes (Pound \& Rebka 1960). However, it has been argued
that if a quasar of moderate redshift has a mass near those of galaxies
(say, 10$^{11}$ times that of the Sun), its radius will be a few percent of
a light-year. Thus gravitational attraction is large and ``gravitational
implosions'' may occur. Then the stability of such an object is questionable
(Greenstein \& Schmidt 1964). Nevertheless, the possibility that a quasar is
much more massive than a galaxy should not be excluded. Correspondingly, a
quasar of moderate redshift can have a radius much larger than a light-year
as indicated in some observations and then its density is much lower. They
together lead to a much weaker gravitational force. Thus the stability may
not be a problem. In this investigation we will reexamine the gravitational
redshift as an intrinsic redshift for quasars. However, the gravitational
redshift is based on a new theory which is fundamentally different and may
lead to different consequences particularly for high-redshift cases with
strong gravitational potentials.

Specifically, it is proposed that the gaseous material constituting a quasar
forms a gravitational potential which in turn will be shown to lower the
energies of quantum states and hence transition frequencies of atoms of the
constituent gas. The total mass of the quasar contributes to this potential.
However, based on the local-ether theory discussed in the next section, it
is supposed that when a quasar breaks into two or more well separated
pieces, then the constituent material of each fragment contributes to the
formation of an individual gravitational potential associated with this
fragment. And the quantum states and transition frequencies are determined
solely by one of the gravitational potentials, rather than by all of them.
Moreover, as part of the quasar forms stars, the constituent material of
each star contributes to the formation of an individual gravitational
potential associated with this star, but no longer to the potential of the
parent quasar. Thus, during the evolution process with quasar fragmentation
and star formation, the masses of quasars decrease and hence their
gravitational potentials become weaker. Consequently, the transition
frequencies become lower and lower with the evolution process. After the
star formation is completed, a quasar evolves into ordinary galaxies or
clusters of galaxies with low redshift and high luminosity. Thus the
gravitation-induced redshift in conjunction with the ejection model and the
local-ether theory is used to account for the observed variation of redshift
in quasars. Quantitatively, based on the fragmentation of quasars, a series
of preferred values of redshift are derived. They will be compared with the
Karlsson peaks in the redshift distributions. The dependence of transition
frequency on the gravitational potential in turn will be derived from a
proposed wave equation based on the local-ether theory, as discussed in the
following sections. From the local-ether wave equation, various consequences
have been derived to account for a wide variety of phenomena related to
special relativity, general relativity, electromagnetics, and quantum
mechanics (Su 2005). The proposed intrinsic redshift then presents an
additional consequence of this wave equation.

\vspace{1cm}

\noindent {\large \textbf{2. Local-Ether Theory}}\\[0.2cm]
It has been proposed that in the region under a sufficient influence of the
gravitation due to the Earth, the Sun, or another celestial body, there
forms a local ether which in turn is stationary with respect to the
gravitational potential of the respective body (Su 2001b, Su 2005). Thereby,
each local ether together with the gravitational potential moves with the
associated celestial body. According to this model, the earth local ether is
inside the sun local ether, which in turn is inside the galaxy local ether,
and so on. The earth local ether together with earth's gravitational
potential moves with earth's orbital motion around the Sun, but not with
earth's rotation. Thus the earth local ether is stationary in an
earth-centered inertial (ECI) frame, while the sun local ether is stationary
in a heliocentric inertial frame. The earth local ether should be at least
so large as to encompass geostationary satellites.

The local-ether theory has been used to account for the Sagnac effect due to
earth's rotation and the null effect of earth's orbital motion in the
propagation of earthbound electromagnetic waves in GPS and the
intercontinental microwave link, where the Sagnac effect is related with a
modification of propagation time for electromagnetic waves due to motion of
the receiver. Meanwhile, in the interplanetary radar, it accounts for the
Sagnac effect due to both of the rotational and orbital motions of the Earth
around the Sun. It also accounts for the null effect of orbital motion of
the Sun in the earthbound and interplanetary propagations. Furthermore, the
local-ether propagation model has been used to account for the apparent null
effect in the Michelson-Morley experiment.

Based on this brand-new theory, the local ethers associated with their
respective planets, stars, galaxies or quasars, are of finite extent and are
supposed to form a hierarchy structure. Thus the local ether of a planet is
immersed locally inside the local ether of the associated star, which in
turn is inside the local ether of the associated galaxy or quasar. However,
the various local ethers are \textit{exclusive}. At a certain position, only
the lowest-level local ether determines propagation of waves or properties
of particles located there. Thereby, it is inferred that the gravitational
potential of the sun has no effects on the phenomena associated entirely
with earthbound waves or particles. This is in accord with the situation
that the gravity at a given position or the tick rate of an atomic clock
located at a given position on or near the ground is identical between at 
\textrm{noon} and \textrm{midnight}, though the distance of the position
from the Sun is different at different hours of a day. And the rate of
atomic clocks onboard various GPS satellites remains unchanged while they
are orbiting in circular orbits around the Earth and thus their distances
from the Sun are varying.

A quasar is supposed to be composed of a huge gas cloud of atoms, plasma,
and dust. In the region under a sufficient influence of the gravitational
potential of the quasar, there forms a local ether associated with this
quasar. However, due to some internal mechanisms like nonuniformity in the
distribution of particle velocity or density, the gas cloud may fragment
into two or more clouds of smaller sizes. When the fragments of the quasar
are well separated from each other, the local ether and gravitational
potential associated with the previous quasar are supposed to split by some
mechanism into multiple individual local ethers and gravitational potentials
associated with their respective fragments. Based on the local-ether theory,
the quantum states and transition frequencies of atoms, molecules, or ions
of the gas constituting a quasar or fragmented quasar which forms a local
ether are determined by the gravitational potential associated with this
local ether, but not by the potentials with other local ethers even in close
proximity. As the mass and gravitational potential of a fragmented quasar
are smaller than those of the previous quasar, the redshift becomes lower
and lower after repeated fragmentation.

In the meantime, by gravitational attraction of gas, various stars are
formed gradually in each quasar or fragmented quasar. Then various
individual local ethers and gravitational potentials associated with their
respective stars are formed inside the quasar local ether, which in turn
covers the whole region of the quasar but excludes the domains of the
numerous stellar local ethers enclosed by it. The matter of a star
contributes to the gravitational potential and local ether associated with
the star and affects properties of atoms located within this ether; however,
it no longer contributes to the upper-level gravitational potential and
local ether associated with the surrounding parent quasar. Consequently, the
gravitational potential associated with the parent quasar and hence its
redshift decrease gradually with star formation. On the other hand, the
redshifts of emission from the stars are expected to be very low, as the
gravitational potentials of individual stars are very weak. Thus, by
fragmentation and condensation, a quasar eventually evolves into galaxies or
cluster of galaxies surrounded by some dilute intergalactic gas. During this
evolution process, the redshifts of all the objects developed from a quasar
tend to decrease with time if they are not low.

\vspace{1cm}

\noindent {\large \textbf{3. Ejection Model for Quasars and its Evidence}}\\[%
0.2cm]
It has been doubted for a long time whether the distances of quasars are
necessarily as far as what their high redshifts indicate, as quasars of high
redshifts seem to be physically associated with a nearby active galaxy of a
low redshift. The most compelling mechanism for this physical association
could be the ejection model which has been proposed in as early as 1967 and
states that some material may be ejected in opposite directions from a
central galaxy of low redshift and high luminosity and the ejected material
is responsible for the formation of other galaxies or quasars (Arp 1967).
The ejection may involve an accretion disk and has a tendency to along the
rotation axis of the parent galaxy. By virtue of conservation of momentum
with the ejection, quasars tend to form on both sides of that axis. Then,
near the center galaxy, a pair of quasars are formed from the clouds of gas
associated with the ejection. By virtue of the momentum gained from
ejection, quasars gradually move away from the center galaxy. Meanwhile,
quasars may condense due to gravitation and then star formation begins. On
the other hand, due to internal disturbance or to nonuniformity in particle
velocity and density, quasars tend to grow in size and may even fragment
into pieces of smaller sizes. The quasar fragmentation is in accord with the
observation that ``\textrm{radio} and x-ray sources ejected from active
galaxies are often double or triple'' (Arp \& Russell 2001). Thus quasars
evolve into quasar-like galaxies and then into normal galaxies or clusters
of galaxies. After the quasars moved away, further ejection from the parent
galaxy forms a newborn pair of quasars. And the evolution process repeats
itself. Thus quasars, quasar-like galaxies, and then normal galaxies will be
distributed in that order away from the parent galaxy along the ejection
path on both sides of the parent galaxy.

Some observations even indicate that there remains a trail or wake extending
out from the active galaxy toward the quasars. It is seen from the \textit{%
HST} image given in (Galianni et al. 2005) that a quasar of strong x-ray
source seems to lie within NGC 7319, one of Stephan's Quintet group of the
Seyfert galaxies, and is located only 8 arcsec from the center of the
galaxy. And a \textsf{V}-shaped filament extends by a few arcsec from the
nucleus of the galaxy toward the quasar (Galianni et al. 2005). Further,
there can be some connection or bridge between the active galaxy and the
nearby quasar. The physical association between these objects, instead of a
fortuitous projection, can be revealed from such a connection. The galaxy
Arp 220 of $z=0.018$ has a group of companion galaxies of $z\sim 0.09$ and
located as close as about 2 arcmin to it. Further, from H I contours it has
been observed that a stream of hydrogen is drawn out of the parent galaxy
Arp 220 and ends exactly on that companion group (Arp 2001). For the active
galaxy NGC 3628 of $z=0.0028$, it is seen from x-ray contours that a
filament extends from the nucleus of the galaxy and ends on two nearby
quasars of $z=0.995$ and 2.15, respectively (Arp et al. 2002). And it has
been observed from optical images that a filament is situated along the line
connecting the compact objects of NEQ3 which are of high redshifts and the
main galaxy (Guti\textrm{\'{e}}rrez \& L\textrm{\'{o}}pez-Corredoira 2004).
These trails or bridges provide quite direct evidence for the ejection model.

Furthermore, we present some arguments to support the ejection model. Based
on the local-ether theory, the material constituting a quasar forms a
gravitational potential associated with the quasar, which in turn is
determined by the size and density of the cloud of ejected gas. When a
quasar breaks due to internal nonuniformity or disturbance, the split
gravitational potentials become weaker. When a quasar grows in size due to
velocity nonuniformity, its gravitational potential becomes weaker. And when
part of a quasar forms stars, its gravitational potential also becomes
weaker. These make the redshifts of quasars decline, while the starburst
makes their luminosities stronger. This contrast is more pronounced when the
star formation is maturer. Thereby, the ejection momentum and the evolution
process with quasar fragmentation and star formation lead to the consequence
that the center galaxy is of the lowest redshift and the highest luminosity,
the nearby quasar is of the highest redshift and the lowest luminosity, and
of the companion quasars or galaxies the redshifts decrease and the
luminosities increase with their distances away from the center galaxy.

This trend is in accord with a configuration proposed in the literature and
with many observations related to various center galaxies, such as the
low-redshift galaxies M82, M101, NGC 6217, and NGC 470/474, to name just a
few (Arp \& Russell 2001). As an example, it has been reported that there
are two galaxy clusters A873 and A910 which have very close redshifts and
are very well aligned and fairly equally spaced across the very bright and
very active galaxy M82. Along the line connecting A873 and A910, there are
four rather bright quasars. And ``\textrm{close} to M82 a dense group of
quasars was serendipitously discovered along a line slightly rotated from
its minor-axis direction of ejection'' (Arp \& Russell 2001). The ejection
path has a tendency to be along the rotation axis and the ejection in
transverse directions is expected to encounter resistance. However, the
ejection with multiple directions seems to be demonstrated from a\textbf{\ }%
tight group of galaxies of x-ray sources close to the low-redshift central
galaxy NGC 383, as most of them ``\textrm{lie} on opposite ends of diameters
passing close to the central galaxy'' (Arp 2001). In some cases the ejection
path can be in an $\mathsf{S}$ or spiral shape, such as the structure with
the low-redshift parent galaxy Arp 220 (Arp 2001). This may be due to
rotation of the ejection axis itself.

We next go on to consider the connection among the fragments. A quasar may
break into two or more pieces of smaller sizes. When they are well separated
from each other, an individual gravitational potential will establish for
each fragment. As their sizes can be similar or different, their respective
gravitational potentials and hence the redshifts can also be so. This
situation of a group of sources in close proximity with similar or different
redshifts has been observed in many cases. A remarkable one is NEQ3 which is
composed of three compact objects which in turn look like to be connected
together without intermediate spaces (see Figures 1 and 2 in Guti\textrm{%
\'{e}}rrez \& L\textrm{\'{o}}pez-Corredoira 2004) and have redshifts of $%
z=0.1935$, 0.1939, and 0.2229, respectively. They are believed to be
physically associated with respect to each other, because of the close
proximity and of the observation that ``\textrm{the} main halo seems to
surround the three objects uniformly'' (Guti\textrm{\'{e}}rrez \& L\textrm{%
\'{o}}pez-Corredoira 2004).

Further, we consider the transition of redshift for the emission from the
regions across the space between two nearby objects which may be due to
ejection or fragmentation and have dissimilar redshifts. If the emission is
determined by both of the gravitational potentials, it can be expected that
the redshift gradually changes across the intermediate space. However, based
on the local-ether theory, the gravitational potential governing the
emission is exclusive and unique and hence the redshift is determined solely
by either of the two potentials. Specifically, consider the H$\alpha $
emission observed in MCG 7-25-46 which is a system with two galaxies of
different redshifts, $z=0.003$ for the main galaxy and $z=0.098$ for the
minor one. The uniqueness of the gravitational potential is in accord with
the observation that ``\textrm{the} H$\alpha $ emission at $z=0.003$
finishes exactly where the H$\alpha $ emission at $z=0.098$ begins and there
is no overlap in the two emissions'' (L\textrm{\'{o}}pez-Corredoira \& Guti%
\textrm{\'{e}}rrez 2005). This kind of abrupt change was also observed for
NGC 7320 and the nearby galaxies (Guti\textrm{\'{e}}rrez et al. 2002). The
phenomenon of abrupt change in redshift without overlap or smooth transition
provides further support for the local-ether theory in conjunction with the
ejection model, though it may be viewed instead as evidence against the
connection between the nearby objects and for the projection effect (Guti%
\textrm{\'{e}}rrez et al. 2002). Thus the local-ether theory provides
physical origins of the intrinsic redshift, its aging, and of its spatial
variation. In the following sections a quantitative treatment of the
redshift will be given.

\vspace{1cm}

\noindent {\large \textbf{4. Local-Ether Wave Equation and Gravitational
Redshift}}\\[0.2cm]
Under the influence of the gravitational potential $\Phi _{g}$ due to a
celestial body and the electric scalar potential $\Phi $ due to a charged
particle, it is postulated that the matter wave $\Psi $ associated with a
particle of charge $q$ is governed by the local-ether wave equation proposed
to be 
$$
\left\{ \nabla ^{2}-\frac{n_{g}}{c^{2}}\frac{\partial ^{2}}{\partial t^{2}}%
\right\} \Psi (\mathbf{r},t)=\frac{\omega _{0}^{2}}{c^{2}}\left\{ 1+\frac{2}{%
\hbar \omega _{0}}q\Phi (\mathbf{r},t)\right\} \Psi (\mathbf{r},t),\eqno
(1) 
$$
where $\hbar $ is Planck's constant divided by $2\pi $, the natural
frequency $\omega _{0}$ as well as the charge $q$ is supposed to be an
inherent constant of the effector particle, and the position vector $\mathbf{%
r}$ and the time derivative are supposed to referred specifically to the
associated local-ether frame. The gravitational index $n_{g}$ is defined as 
$$
n_{g}(\mathbf{r})=1+\frac{2}{c^{2}}\Phi _{g}(\mathbf{r}),\eqno
(2) 
$$
where the gravitational potential is given in a positive-definite form and
the gravitational index is always greater than unity.

Based on the local-ether wave equation, a time evolution equation similar to
Schr\textrm{\"{o}}\-ding\-er's equation can be derived. Then, by evaluating
the velocity and then the acceleration in a quantum-mechanical approach, it
has been shown that the gravitational force and the electrostatic force
exerted on the charged particle is given by (Su 2002a, Su 2005) 
$$
\mathbf{F}=-q\nabla \Phi +m_{0}\nabla \Phi _{g}.\eqno
(3)
$$
Moreover, the gravitational mass associated with the gravitational force and
the inertial mass with the electromagnetic force have been shown to be
identical to the natural frequency by the familiar form 
$$
m_{0}=\frac{\hbar }{c^{2}}\omega _{0}.\eqno
(4)
$$
In addition to the electrostatic force, the other electromagnetic forces
have also been derived from the wave equation with a refinement (Su 2002b,
Su 2005). Thereby, the local-ether wave equation leads to a unified quantum
theory of electromagnetic and gravitational forces in conjunction with the
origin and the identity of inertial and gravitational mass. (The wave
equation proposed in (Su 2002a, Su 2005) is given in a slightly different
form. However, this difference makes the results unchanged.) The wave
equation can also be applied to electromagnetic wave, of which the natural
frequency is supposed to be zero. It has been shown that, with the
modification where the index $n_{g}$ in (1) is replaced by its square, the
wave equation for electromagnetic wave has an immediate consequence that the
wave propagates at a reduced speed of $c/n_{g}$, which in turn leads to the
familiar phenomena of the gravitational deflection of light by the Sun and
the increment of echo time in the interplanetary radar (Su 2001a, Su 2005).

We then go on to discuss another consequence of the wave equation for the
matter wave that is bound in an atom or a molecule. It is known that the
wave then exists in one of some particular quantum states in which the
temporal variation of $\Psi $ can follow the one of a pure time harmonic $%
e^{-i\omega t}$, as a consequence of resonance. That is, $\Psi (\mathbf{r}%
,t)=\psi (\mathbf{r})e^{-i\omega t}$, where $\psi $ is independent of time
when observed in the local-ether frame. For such a wave, a manipulation
associated with the expectation value renders the preceding wave equation
into the algebraic relation 
$$
\omega ^{2}=\frac{1}{n_{g}}\omega _{0}^{2}\left\{ 1+\frac{2}{\hbar \omega
_{0}}\left\langle q\Phi \right\rangle -\frac{c^{2}}{\omega _{0}^{2}}%
\left\langle \nabla ^{2}\right\rangle \right\} ,\eqno
(5)
$$
where the wavefunction is supposed to be normalized. As seen from the wave
equation (1), the angular frequency $\omega $ of a bound matter wave depends
on the gravitational potential, while the wavefunction $\psi $ is
independent of the potential. Thus the angular frequency can be easily
evaluated from the preceding relation if the wavefunction in the absence of
the gravitational potential is known.

Ordinarily, the scalar potential $\Phi $ and the spatial variation of $\Psi $
are weak. Thus, by evaluating the square root of the right-hand side of the
preceding frequency formula with the binomial expansion to the first order,
the quantum energy $\hbar \omega $ of the matter wave bound in an atom can
be given by 
$$
\hbar \omega =\frac{1}{\sqrt{n_{g}}}\left\{ \hbar \omega _{0}+\left\langle
q\Phi \right\rangle -\frac{\hbar ^{2}}{2m_{0}}\left\langle \nabla
^{2}\right\rangle \right\} .\eqno
(6) 
$$
It is seen that the gravitational potential, the electric scalar potential,
and the spatial variation of the wavefunction modify the quantum energy. As
in quantum mechanics, the frequency of the light emitted from or absorbed by
an atom or a molecule is supposed to be given by the transition frequency
which in turn is associated with the difference in energy between the two
quantum states participating in the transition. It is noted that the major
term of the quantum energy, namely, the first term on the right-hand side of
the preceding energy formula is identical in different states and hence its
effect on the transition frequency cancels out. Thus the state transition
frequency is due to the minor energy terms.

For the cases with a weak gravitational potential, the transition frequency
can be given in the familiar form (Su 2001c, Su 2005) 
$$
f=f_{0}\left( 1-\frac{\Phi _{g}}{c^{2}}\right) ,\eqno
(7) 
$$
where $f_{0}$ denotes the transition frequency in the absence of the
gravitational potential. This gravitation-induced decrease of frequency is
known as the gravitational redshift and has been demonstrated in the
frequency deviation in the Pound-Rebka experiment dealing with emission and
absorption of gamma ray, in the clock-rate difference in the Hafele-Keating
experiment with cesium atomic clocks under circumnavigation, and in the
clock-rate adjustment in atomic clocks onboard GPS satellites before their
launch to circular orbits. In the Pound-Rebka experiment the variation of
frequency is commonly attributed to the influence of gravitation on the
photons which are traveling through a region where the gravitational
potential varies spatially. However, based on the local-ether wave equation,
the variation of frequency is due to the gravitation-induced decrease of
quantum-state energy and transition frequency. After emission from atoms,
the frequency of the electromagnetic wave will no longer change. (However,
the observed frequency is still subject to change by virtue of the Doppler
effect.) Based on this, we have presented reinterpretations for the
aforementioned experiments (Su 2005). Furthermore, the local-ether wave
equation leads to the consequence that the quantum-state energies of an atom
decrease with its speed $v$ by the factor $\sqrt{1-v^{2}/c^{2}}$, which in
turn looks like the famous factor adopted in the Lorentz transformation of
space and time. The derived speed-dependent transition frequency has been
used to account for the east-west directional anisotropy in the
Hafele-Keating experiment and for another factor of the clock-rate
adjustment in GPS (Su 2001c, Su 2005). However, this effect is not so
important in high redshift, unless the atom speed is high enough, and is not
considered further in this investigation.

For the general case the frequency formula (6) yields that the
gravitation-dependent transition frequency of the atoms placed on a
celestial body with a gravitational potential $\Phi _{g}$ is given by 
$$
f=f_{0}/\sqrt{1+2\Phi _{g}/c^{2}}.\eqno
(8) 
$$
The corresponding gravitation-induced intrinsic redshift is then given by 
$$
z=\sqrt{1+2\Phi _{g}/c^{2}}-1.\eqno
(9) 
$$
Based on this formula, a quantitative analysis of the intrinsic redshift,
its variation, and of its preferred values will be given in the following
section.

\vspace{1cm}

\noindent {\large \textbf{5. Estimates of Gravitation-Induced Intrinsic
Redshift}}\newline
{\large \textbf{and Preferred Values}}\\[0.2cm]
As in classical gravitation, the gravitational potential associated with a
celestial object is supposed to be determined by the total mass $M$ of the
constituent material by $\Phi _{g}=GM/R$, where $G$ is the gravitational
constant and $R$ is the separation distance from the center of the object.
In an attempt to estimate this potential, one should know the size and
density of the object. According to the ejection model, quasars are formed
from gas of atoms, plasma, and dust ejected from a parent active galaxy. The
optical spectra from quasars or quasar-like objects near M82, which is the
nearest active galaxy to our Milky Way galaxy, reveal that the constituent
atoms or ions include hydrogen, helium, carbon, nitrogen, oxygen, neon,
magnesium, silicon, sulfur, and iron (Burbidge et al. 2003).

From the \textit{HST} image given in (Galianni et al. 2005), the angular
diameter of the quasar quite close to the nucleus of NGC 7319 is estimated
to be 0.7 arcsec. It is widely accepted that the distance of NGC 7319 from
the Earth is about 300 million light-years (Wikipedia). If this quasar is
indeed physically close to the galaxies and of that distance, then the
physical diameter of this quasar will be about 1000 light-years. As a
comparison, the diameter of the Milky Way is 10$^{5}$ light-years. Suppose
that the gas cloud forming that quasar has a spherical shape of that
diameter. Next, we estimate its density. For comparison, the average density
of the Sun is about $10^{3}$ kg/m$^{3}$ and the density of dry air at
standard ambient temperature and pressure is $1.2$ kg/m$^{3}$. Interstellar
gas is more tenuous. For molecular clouds in an interstellar space, the
concentration varies from 10$^{9}$ to 10$^{12}$ atoms per cubic meter
(Wikipedia). This amounts to a density ranging from $3.4\times 10^{-18}$ to $%
3.4\times 10^{-15}$ kg/m$^{3}$, as the particles are molecular hydrogen.
Optical spectra indicate that quasars contain heavier atoms or ions. They
probably contain dust as well. Anyway, a quasar should be denser than an
interstellar gas. This is because the star formation is difficult to
initiate without a denser gas and, on the other hand, most of the gas will
be depleted after the completion of star formation. Suppose the quasar close
to NGC 7319 has a density of $6\times 10^{-11}$ kg/m$^{3}$. Owing to its
size, the total mass of the quasar will be as large as $2.7\times 10^{46}$
kg. For comparison, the mass of the Sun and of the Milky Way are $2\times
10^{30}$ and $\sim 10^{42}$ kg, respectively. It is noted that for such a
massive quasar the Schwarzschild radius, given by $R_{s}=2GM/c^{2}$, is much
greater than its radius, while its density is very low. Based on the
local-ether theory, a Schwarzschild radius being greater than the radius of
an object implies that on the surface of the object, the gravitational
potential is so strong that $\Phi _{g}>c^{2}/2$ and hence the
gravitation-induced intrinsic redshift $z>0.414$.

Then suppose that the material of the gas cloud together forms a local ether
associated with the quasar. Thus the gravitational potential on the surface
of the quasar, when normalized to $c^{2}$, is $\Phi _{g}/c^{2}=4.2$.
Although this potential is extraordinarily strong, the gravity of
acceleration on the surface of such a massive quasar is less than one
percent of that on the surface of the Earth. Then, according to the redshift
formula, the emission from or absorption by atoms or ions placed near the
surface will have a gravitational redshift of $z=2.07$. This value is close
to the observed redshift of $z=2.114$. A better agreement can be reached
simply with a slight adjustment in the size or density. On the other hand,
suppose another quasar has a lower redshift of $z=0.06$ and a larger
diameter of 3000 light-years. Thus the normalized gravitational potential $%
\Phi _{g}/c^{2}\sim 0.06$ and the density is as low as $\sim 10^{-13}$ kg/m$%
^{3}$. For comparison, the normalized gravitational potential on the surface
of the Earth is about $7\times 10^{-10}$ and that on the surface of the Sun
is about $2\times 10^{-6}$. For the two cases in the solar system, the
gravitational redshift is very small. Thus, based on the ejection model and
the local-ether theory, the wide variation in redshift can be ascribed to a
variation in density and size of the gas cloud, which in turn can be due to
the strength of initial ejection from the parent galaxy, to the speed at
which the cloud moves away from the galaxy, to the gas expansion, to the
fragmentation of gas clouds, and to the star formation.

As the size and density tend to vary widely, it seems that redshifts of
quasars vary in a random way. However, from an analysis of about 600 quasars
it has been found that the redshifts tend to have some preferred values and
thus the distribution of the redshifts exhibits some preferred peaks
(Karlsson 1977). Further, these peaks were found to correspond to a
geometric series in $(1+z)$. That is, the peaks are related by the Karlsson
formula (Karlsson 1971) 
$$
1+z_{n}=(1+z_{0})1.227^{-n},\eqno
(10) 
$$
where the index $n$ is an integer. Thus the quantities ln$(1+z_{n})$ have a
periodicity of 0.205. Statistically, quasars with redshifts close to $z=1.95$
have been found to be quite prominent in their observed amount (Burbidge \&
Burbidge 1967). And quasars with redshifts much higher than 2 are rarely
found. However, the preceding formula can be used to trace the preferred
peaks back to high-redshift cases by letting the index $n$ be negative.
Thus, by adopting the preferred value of 1.956 as the zeroth redshift
(Karlsson 1971), the Karlsson formula yields that the redshifts are peaked
at $z=5.70$, 4.46, 3.45, 2.63, 1.96, 1.41, 0.96, 0.60, 0.30, and 0.06 with $%
n=-4$ to 5. The Karlsson peaks are also confirmed in several subsequent
investigations (Barnothy \& Barnothy 1976, Arp et al. 1990, Karlsson 1990,
Burbidge \& Napier 2001, Burbidge 2003, Napier \& Burbidge 2003),
particularly for the structures with multiple quasars around a nearby galaxy
of low redshift and high luminosity (Arp et al. 1990, Karlsson 1990,
Burbidge \& Napier 2001). But it seems that no physical interpretation for
the periodicity has been proposed. In what follows, we present a model for
the preferred peaks based on the local-ether theory in conjunction with the
ejection model.

Due to nonuniformity in particle velocity and density or to some internal
disturbance, a quasar may break into pieces of smaller sizes. By reason of
symmetry, it seems to have a good chance to break into two pieces of
identical or similar sizes. Suppose the fragments are also spherical and the
density remains unchanged. Thereby, their radius is shorter than the
previous one by a factor of $2^{-1/3}$ and the gravitational potential on
the surface of either fragment will decrease by a factor of $2^{-2/3}$. A\
fragment may even break again, and again. Thus, after the $n$th splitting in
half, the gravitation-induced intrinsic redshift of one such fragment is
given by 
$$
1+z_{n}=\sqrt{1+[(1+z_{0})^{2}-1]2^{-2n/3}},\eqno
(11)
$$
where the quantity $(1+z_{0})^{2}-1$ denotes two times the normalized
gravitational potential corresponding to the zeroth redshift $z_{0}$. For
the cases of very high redshifts, the preceding formula can be approximated
as 
$$
1+z_{n}\simeq (1+z_{0})1.26^{-n}.\eqno
(12)
$$
By adopting the preferred value of 1.956 as the zeroth redshift again,
formula (11) leads to the prediction that the preferred intrinsic redshifts
are $z=6.08$, 4.65, 3.53, 2.64, 1.96, 1.42, 1.02, \textrm{0.71}, \textrm{0.49%
}, 0.33, \textrm{0.22, 0.14, 0.09}, and 0.06 with $n=-4$ to $9$.

For $n=-1$ to 2, the predicted values are close to the corresponding
Karlsson peaks. For some of the histograms given in (Karlsson 1977, Arp et
al. 1990, Burbidge \& Napier 2001), our results can fit a little better. The
Karlsson peak of 0.60 is quite disparate from our predicted values. This
peak happens to be the mean between the predicted values of 0.71 and 0.49. A
redshift distribution around 0.6 may actually be a merger of two close
distributions around 0.71 and 0.49, respectively. This kind of merging due
to closeness seems to find preliminary support from some of the histograms
given in (Burbidge \& Napier 2001, Burbidge 2003). New preferred values of
0.22, 0.14, and 0.09 are also predicted. Redshifts close to these values can
be found in the literature. However, low redshift peaks are expected to be
smeared, since other affecting factors of uneven splitting, gas expansion,
and star formation, will accumulate with time and then they together with
the Doppler effect and the speed-dependent transition frequency will become
comparatively significant for low-redshift cases. On the other hand, for the
peaks at higher redshifts with $n=-2$ to $-4$, our predicted values are a
few percent greater than the corresponding Karlsson peaks. Although quasars
of such high redshifts are rarely found, this discrepancy as well as the
merging provides a means to test the proposed mechanism based on the
repeated even fragmentation and the gravitation-induced intrinsic redshift.

\vspace{1cm}

\noindent {\large \textbf{6. Conclusion}}\\[0.2cm]
According to the ejection model, quasars of high redshift and low luminosity
originate from the ejected gas from an active galaxy of low redshift and
high luminosity. Then quasars move away from the parent galaxy and
eventually evolve into galaxies. Meanwhile, further ejection forms newborn
quasars, which in turn are accompanied by the quasars or galaxies formed
earlier. From observations these companion quasars or galaxies tend to have
medium redshift and luminosity, with an extent depending on their distances
from the center galaxy.

Based on the local-ether wave equation, it is shown that the quantum-state
energies and transition frequencies of atoms or ions placed in a celestial
object decrease under the influence of the associated gravitational
potential. Thus the gravitational redshift is proposed as an intrinsic
redshift associated with the emission from stars, galaxies, or quasars.
Initially, quasars are massive and their respective gravitational potentials
are strong. Thus the redshifts of quasars are high. However, in the
evolution process with quasar fragmentation and star formation, the masses
of quasars decrease and their gravitational potentials become weaker. Thus
their redshifts become lower and lower, while the starburst makes their
luminosities stronger. Thereby, the proposed gravitation-induced intrinsic
redshift in conjunction with the ejection model is in accord with the
observed variations of redshift and luminosity among the parent galaxy,
nearby quasars, and the companion quasars and galaxies. The local-ether
theory is further supported by the abrupt change in redshift without overlap
or smooth transition observed in MCG 7-25-46 and NGC 7320, though this
phenomenon may be viewed instead as evidence for the projection effect.

Quantitatively, based on the observed angular diameter of the high-redshift
quasar near NGC 7319 and on the density of interstellar gas, we estimate the
physical diameter and density of the quasar. Thereby, the calculated
gravitational redshift can agree with the observed high redshift, while the
corresponding gravitational force is weak. Further, based on the repeated
fragmentation of quasars in half, a series of preferred values of redshift
are predicted. Most of them are close to the corresponding Karlsson peaks,
except that two new preferred values of 0.71 and 0.49 are predicted, but our
results lack the Karlsson peak of 0.60. It is expected that a distribution
around the last peak may be a merger of two close distributions around the
former two, respectively. This merging then provides a quantitative means to
test the proposed mechanism based on the quasar fragmentation and the
local-ether theory.

\vspace{2cm}

\noindent {\large \textbf{References}}

$\ $

\noindent Arp, H. 1967, ApJ, 148, 321

\noindent \underline{$\hspace{1.2cm}$}. 2001, ApJ, 549, 780

\noindent Arp, H., Bi, H.G., Chu, Y., \& Zhu, X. 1990, A\&A, 239, 33

\noindent Arp, H., Burbidge, E.M.,Chu, Y., Flesch, E., Patat, F., \&
Rupprecht, G. 2002, A\&A, 391,\linebreak $\hspace*{0.5cm}$833

\noindent Arp, H., \& Russell, D. 2001, ApJ, 549, 802

\noindent Barnothy, J.M., \& Barnothy, M.F. 1976, PASP, 88, 837

\noindent Burbidge, E.M., Burbidge, G., Arp, H.C., \& Zibetti, S. 2003, ApJ,
591, 690

\noindent Burbidge, G.R. 2003, ApJ, 585, 112

\noindent Burbidge, G.R., \& Burbidge, E.M. 1967, ApJ, 148, L107

\noindent Burbidge, G., \& Napier, W.M. 2001, AJ, 121, 21

\noindent Galianni, P., Burbidge, E.M., Arp, H., Junkkarinen, V., Burbidge,
G., \& Zibetti, S. 2005,\linebreak $\hspace*{0.5cm}$ApJ, 620, 88

\noindent Greenstein, J. \& Schmidt, M. 1964, ApJ, 140, 1

\noindent Guti\textrm{\'{e}}rrez, C.M., \& L\textrm{\'{o}}pez-Corredoira, M.
2004, ApJ, 605, L5

\noindent Guti\textrm{\'{e}}rrez, C.M., L\textrm{\'{o}}pez-Corredoira, M.,
Prada, F., \& Eliche, M.C. 2002, ApJ, 599, 579

\noindent Karlsson, K.G. 1971, A\&A, 13, 333

\noindent \underline{$\hspace{1.2cm}$}. 1977, A\&A, 58, 237

\noindent \underline{$\hspace{1.2cm}$}. 1990, A\&A, 239, 50

\noindent L\textrm{\'{o}}pez-Corredoira, M., \& Guti\textrm{\'{e}}rrez, C.M.
2004, ApJ, 605, L5

\noindent \underline{$\hspace{1.2cm}$}. 2005 (astro-ph/05096302)

\noindent Napier, W. M., \& Burbidge, G. 2003, MNRAS, 342, 601

\noindent Pound, R. V. \& Rebka, Jr., G. A. 1960, \textit{Phys. Rev. Lett}.,
4, 337

\noindent Su, C.C. 2001a, \textit{J. Electromagnetic Waves Applicat.}, 15,
259

\noindent \underline{$\hspace{1.2cm}$}. 2001b,\textit{\ Eur. Phys. J. C},
21, 701

\noindent \underline{$\hspace{1.2cm}$}. 2001c, \textit{Eur. Phys. J. B}, 24,
231

\noindent \underline{$\hspace{1.2cm}$}. 2002a, \textit{J. Electromagnetic
Waves Applicat.}, 16, 40

\noindent \underline{$\hspace{1.2cm}$}. 2002b, \textit{J. Electromagnetic
Waves Applicat.}, 16, 1275

\noindent \underline{$\hspace{1.2cm}$}. 2005, \textit{Quantum
Electromagnetics -- A Local-Ether Wave Equation Unifying Quantum}\linebreak $%
\hspace*{0.5cm}$\textit{Mechanics, Electromagnetics, and Gravitation} (%
\texttt{http://qem.ee.nthu.edu.tw})

\noindent In \textit{Wikipedia}, \texttt{http://wikipedia.org}, arts.
``Stephan's Quintet'', ``Interstellar medium''

$\ $

\end{document}